\documentclass[aps,prb,twocolumn,superscriptaddress]{revtex4-1}

\usepackage{amssymb}
\usepackage{amsfonts}
\usepackage{amsmath}
\usepackage{color}

\usepackage[pdftex]{graphicx}

\begin{document}

\date{\today}

\title{Anisotropy of the coherence length from critical currents in the stoichiometric superconductor LiFeAs }

\author{M. Ko\'{n}czykowski}
\affiliation{Laboratoire des Solides Irradi\'{e}s, CNRS-UMR 7642 \& CEA-DSM-IRAMIS, Ecole Polytechnique, F 91128 Palaiseau cedex, France}

\author{C.~J.~van der Beek}
\affiliation{Laboratoire des Solides Irradi\'{e}s, CNRS-UMR 7642 \& CEA-DSM-IRAMIS, Ecole Polytechnique, F 91128 Palaiseau cedex, France}
\author{M. A. Tanatar}
\affiliation{The Ames Laboratory, Ames, IA 50011, U.S.A.}
\author{V. Mosser}
\affiliation{ITRON/Issy Technology Center, 52 rue Camille Desmoulins, F-92130 Issy-les-Moulineaux, France}

\author{Yoo~Jang~Song}
\affiliation{\mbox{Department of Physics, Sungkyunkwan University, Suwon, Gyeonggi-Do 440-746, Republic of Korea}}

\author{Yong~Seung~Kwon}
\affiliation{\mbox{Department of Physics, Sungkyunkwan University, Suwon, Gyeonggi-Do 440-746, Republic of Korea}}

\author{R.~Prozorov}
%\email{Corresponding author: prozorov@ameslab.gov}
\affiliation{The Ames Laboratory, Ames, IA 50011, U.S.A.}
\affiliation{Department of Physics \& Astronomy, Iowa State University, Ames, IA 50011, U.S.A.}

\date{30 September 2011}

\begin{abstract}
Miniature Hall-probe arrays were used to measure the critical current densities for the three main directions of vortex motion in the stoichiometric LiFeAs superconductor. These correspond to vortex lines along the $c$-axis moving parallel to the $ab$-plane, and to vortex lines in the $ab$--plane moving perpendicular to, and within the plane, respectively. The measurements were carried out in the low-field regime of strong vortex pinning, in which the critical current anisotropy is solely determined by the coherence length anisotropy parameter, $\varepsilon_{\xi}$. This allows for the extraction of $\varepsilon_{\xi}$ at magnetic fields far below the upper critical field $B_{c2}$. We find that increasing the magnetic field decreases the anisotropy of the coherence length.
\end{abstract}

\pacs{74.25-q,74.25.N-,74.25.Sv,74.25.Wx}
\maketitle

The determination of the electronic anisotropy in the superconducting state is a fundamental problem in multi-band type-II superconductors, that has attracted much in interest with the discovery of the iron-based materials.\cite{Johnston2010review} In single band materials with an ellipsoidal Fermi surface, one can describe the anisotropy using  the ratio  $\varepsilon \equiv (m/M)^{1/2} < 1$  of the electron effective masses,  provided that transport along the anisotropy ($c$--) axis of the material is coherent.\cite{Blatter94} This, however, yields an oversimplified picture in which the anisotropy is temperature--independent. In multi-band superconductors, %Therefore, this approach should be generalized. 
the contribution of electronic bands with different, $k-$dependent Fermi velocities and gap values leads to different ratios $\varepsilon_{\lambda}(T) = \lambda_{ab}/ \lambda_{c}$ and  $\varepsilon_{\xi}(T) = \xi_{c}/ \xi_{ab}$ of the in--plane and $c$-axis London penetration depths $\lambda_{ab,c}(T)$ and coherence lengths $\xi_{ab,c}(T)$.  The low temperature value of the penetration depth anisotropy $\varepsilon_{\lambda}(0) = \varepsilon \left( v_{F,c} / v_{F,ab} \right)$ is determined by the anisotropy of the Fermi velocity, while its temperature dependence reflects the relative probabilities of quasi-particle excitation in the two directions. On the other hand, the coherence length  anisotropy $\varepsilon_{\xi} \sim \left(v_{F,c} / v_{F,ab} \right)\left( \Delta_{c} / \Delta_{ab} \right)$ directly depends on the anisotropy of the superconducting gap $\Delta$. As a result of the changing weight of superconductivity on different Fermi surface sheets and  that of intra- and  interband scattering, both $\varepsilon_{\xi}$ and $\varepsilon_{\lambda}$ are temperature\cite{GurevichMgB2review,Kogan2002anisotropy} and field-dependent,\cite{Nojima2006}  behavior exemplified by MgB$_2$,\cite{Nojima2006,Hc2MgB2anisotropy,Fletcher2005MgB2} the iron-based superconductors, \cite{Okazaki2009,Cho2011_LiFeAs_Hc2,Kurita2011Hc2,Khim2011,Song2011EPL,Prozorov2011ROPP} and, possibly, NbSe$_{2}$.\cite{Sonier2005} 

Experimentally, the anisotropy parameter $\varepsilon_{\xi}$ is usually  determined from the ratio of the $c-$axis and $ab-$plane upper critical fields, $B_{c2}^{\parallel c} = \Phi_{0}/2\pi\xi_{ab}^{2}$ and $B_{c2}^{\parallel ab} = \Phi_{0}/2\pi\xi_{ab}\xi_{c}$,\cite{Cho2011_LiFeAs_Hc2,Kurita2011Hc2,Khim2011}  while the ratio of the lower critical fields $B_{c1}^{\parallel c} = (\Phi_{0}/4\pi\mu_{0}\lambda_{ab}^{2})\ln \kappa_{ab}$ and $B_{c1}^{\parallel ab} = (\Phi_{0}/4\pi\mu_{0}\lambda_{ab}\lambda_{c})\ln \kappa_{c}$ is used to evaluate $\varepsilon_{\lambda}$.\cite{Okazaki2009,Song2011EPL} Here, $\Phi_{0} = h/2e$ is the flux quantum, $\kappa_{ab} = \lambda_{ab}/\xi_{ab}$ and $\kappa_{c} = (\lambda_{ab}\lambda_{c}/\xi_{ab}\xi_{c})^{1/2}$. Another approach is the direct measurement of $\lambda$ using differently oriented ac fields.\cite{Prozorov2011ROPP} Hence, $\varepsilon_{\lambda}$ is usually obtained from measurements at low reduced fields $B/B_{c2}$, while $\varepsilon_{\xi}$ is extracted from data in the high field regime close to $B_{c2}$.  %Below, we show that $\varepsilon_{\xi}$ at low fields can be accessed by measuring the critical current density, which depends on $\xi$ since the latter represents the radius of the vortex core.

%More precisely, we determine ... from...
Below, we show that $\varepsilon_{\xi}$ at low fields can be accessed by direct measurements of the critical current density along three principal directions: $j_{ab}^{c}$ for vortex lines along the $c$-axis moving parallel to the $ab$-plane,   $j_{ab}^{ab}$ for vortices parallel to the $ab$--plane and moving parallel to the $c$-axis,  and  $j_{c}^{ab}$ for vortices again parallel to the $ab$--plane, but moving within the plane. Experimentally, this is not a trivial task, as the signal from usual bulk magnetometry for ${\bf B} \parallel ab$ will always involve contributions from both $j_{ab}^{ab}$ and $j_{c}^{ab}$. In Fe-based superconductors, the only work that we are aware of uses transport measurements of the three critical currents in mesoscopic bridges fashioned by focused-ion beam (FIB) lithography in Sm-1111 single crystals.\cite{NatureComm} In what follows, we report on \emph{contactless} measurements using miniature Hall-probe arrays, with the same single crystal positioned in different orientations, which allow one to unambiguously measure the critical current density for the three different situations.

In order to analyze the critical current density, we have rederived known expressions  for the respective cases of weak-\cite{Blatter94} and strong \cite{Ovchinnikov91,vdBeek2002} vortex pinning, for the three relevant magnetic field and current orientations. In doing so, we keep track of $\lambda_{ab,c}(T)$ and $\xi_{ab,c}(T)$ as they appear, combining them into the ratios $\varepsilon_{\lambda}$ and $\varepsilon_{\xi}$ only as a final step.\cite{tbp}  It turns out that  in the regime of strong pinning by extrinsic nm-scale defects, the anisotropy $j_{ab}^{ab}/j_{c}^{ab}$ directly yields $\varepsilon_{\xi}$. In iron-based superconductors, this pinning mechanism is relevant at low magnetic fields.\cite{Kees,Kees1} At intermediate fields,  weak pinning due to scattering by dopant atoms dominates the critical current.\cite{Kees,Kees1} Then $\varepsilon_{\xi}$ is the main (but not the only) contribution to $j_{ab}^{ab}/j_{c}^{ab}$. In order to obtain unambiguous results, one should thus make sure that the critical current is measured in the limit of strong pinning. Thus, we have chosen a superconducting system with reduced intrinsic scattering, in the guise of the (tetragonal) stoichiometric compound LiFeAs.\cite{Wang2008}
%The critical current density is analyzed using a generalized description in which we keep track of the anisotropy contributions of $\lambda(T)$ and $\xi(T)$ as they are, without relating vortex-related parameters to the microscopic expressions that tie these lengths together in a clean single-band approach. It turns out that  in the regime of strong pinning by extrinsic nm-scale defects, relevant for all iron-based superconductors at low magnetic fields,\cite{Kees}  the $j_{ab}^{ab}/j_{c}^{ab}$ anisotropy directly yields $\varepsilon_{\xi}$; in the intermediate-field weak pinning regime\cite{Kees} $\varepsilon_{\xi}$ dominates. In order to exacerbate the contribution of the strong pinning regime, we chose a  superconducting system with reduced intrinsic scattering, in the guise of the (tetragonal) stoichiometric compound LiFeAs.\cite{Wang2008} 
Angle-resolved photoemission \cite{Borisenko2010}, London penetration depth \cite{Kim2011LiFeAsLambda} and first critical field measurements \cite{Song2011EPL} have shown that this is a fully gapped two-band superconductor with moderate anisotropy. One of the cylindrical hole surfaces centered on the $\Gamma$-point has the smaller gap value of $\Delta = 1.5$ meV, while the gap on the more dispersive electron surface around the $M$-point has $\Delta = 2.5$ meV.\cite{Borisenko2010}  Measurements of the anisotropic upper critical field shows that $H_{c2}$ is of mostly orbital character for $H \parallel c-$axis, and Pauli limited for $H \perp c$;\cite{Cho2011_LiFeAs_Hc2,Kurita2011Hc2,Khim2011}  there is evidence for the Fulde-Ferrell-Larkin-Ovchinnikov state for the latter configuration.\cite{Cho2011_LiFeAs_Hc2} A second peak effect (SPE) or ``fishtail'' was reported from magnetization measurements.\cite{Pramanik2010LiFeAs} For $H \parallel c$, the critical current densities range from $\sim 1$\cite{Song2010}  to $\sim 100$ kA/cm$^{2}$.\cite{Pramanik2010LiFeAs} This might be indicative of different defect structures in crystals obtained in different growth procedures. Measurements of the Campbell length on our crystals have shown an even higher ``theoretical'' critical current density of $1 \times 10^{3}$ kA/cm$^{2}$.\cite{Prommapan}

\begin{figure}[t]%
\includegraphics[width=1.35\linewidth]{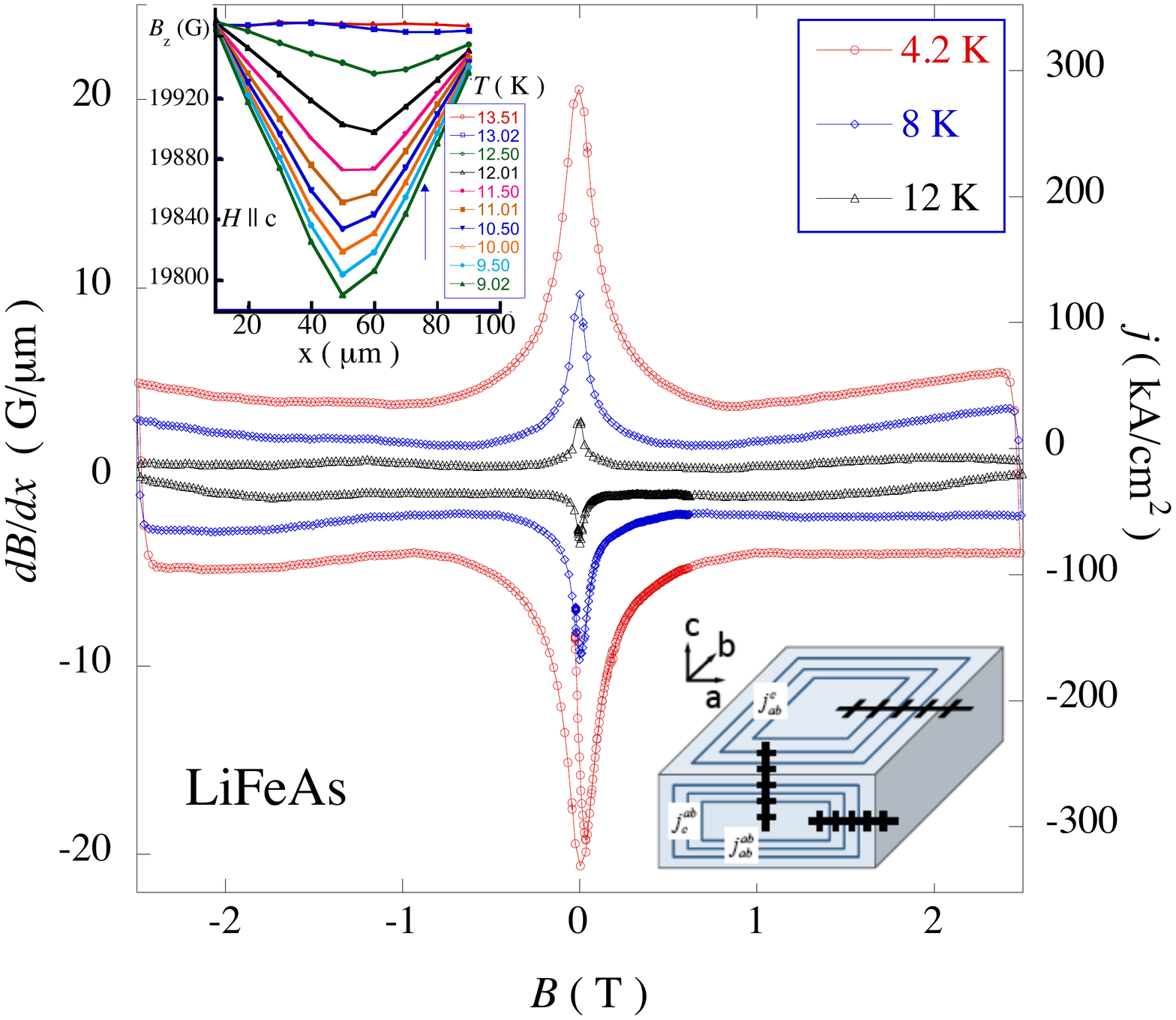}%
\vspace{-7mm}
\caption{(Color online) Lower inset: Experimental scheme, with the three positions of the Hall array (shown as a thick black line with intersecting segments) used to probe the $j_{c}$ for the three possible orientations, as described in the text. Upper inset: Successive profiles of the magnetic induction, obtained on warming after initial zero-field cooling and the application of an external field,  $\mu_{0}H_{a} = 2$ T $\parallel c$. This configuration probes $j_{ab}^{c}$. Main panel: Hysteresis loops of the in--plane  local gradient $dB/dx$ for $\mu_{0}H_{a}\parallel c$. }
\label{fig1}%
\end{figure}

%\section{Experimental details}

Single crystals of LiFeAs were grown in a sealed tungsten crucible using the Bridgman method \cite{Song2011EPL,Song2010} and were transported in sealed ampoules. Immediately after opening, a $0.16 \times 0.19\times 0.480$ mm$^{3}$ rectangular parallelepiped sample was cut with a wire saw, washed and protected in mineral oil. Crystals from the same batch were used for transport as well as AC and DC magnetization measurements. Overall, samples from three different batches
were measured, yielding consistent results. % between them, and with those of Refs. \onlinecite{Song2011EPL,Song2010,Kim2011LiFeAsLambda,Pramanik2010LiFeAS,Cho2011_LiFeAs_Hc2}  
The Hall probe arrays were tailored in a pseudomorphic AlGaAs/InGaAs/GaAs heterostructure using proton implantation. The 10 Hall sensors of the array, spaced by
either 10 or 20 $\mu$m, had an active area of 3 $\times$ 3 $\mu$m$^{2}$, while an 11$^{\mathrm{th}}$ sensor located far from the others was used for the measurement of the applied field. The LiFeAs crystal was positioned appropriately for the measurement of the critical current density in each of the different  orientations, as illustrated in the inset to Fig.~\ref{fig1}. For the measurement of $j_{ab}^{c}$, the crystal  was centered with its $ab$-face on the sensor array, with the array perpendicular to the long edge. For the measurement of $j_{c}^{ab}$ and $j_{ab}^{ab}$, the crystal was centered with its $ac$--face on the array, with the array perpendicular to $c$ and to $ab$, respectively. In all configurations, the local magnetic induction $B$ perpendicular to the Hall sensors (and to the sample surface) was measured along a line across the sample face, in fields up to 2.5~T.

\begin{figure}[t]%
\includegraphics[width=1.35\linewidth]{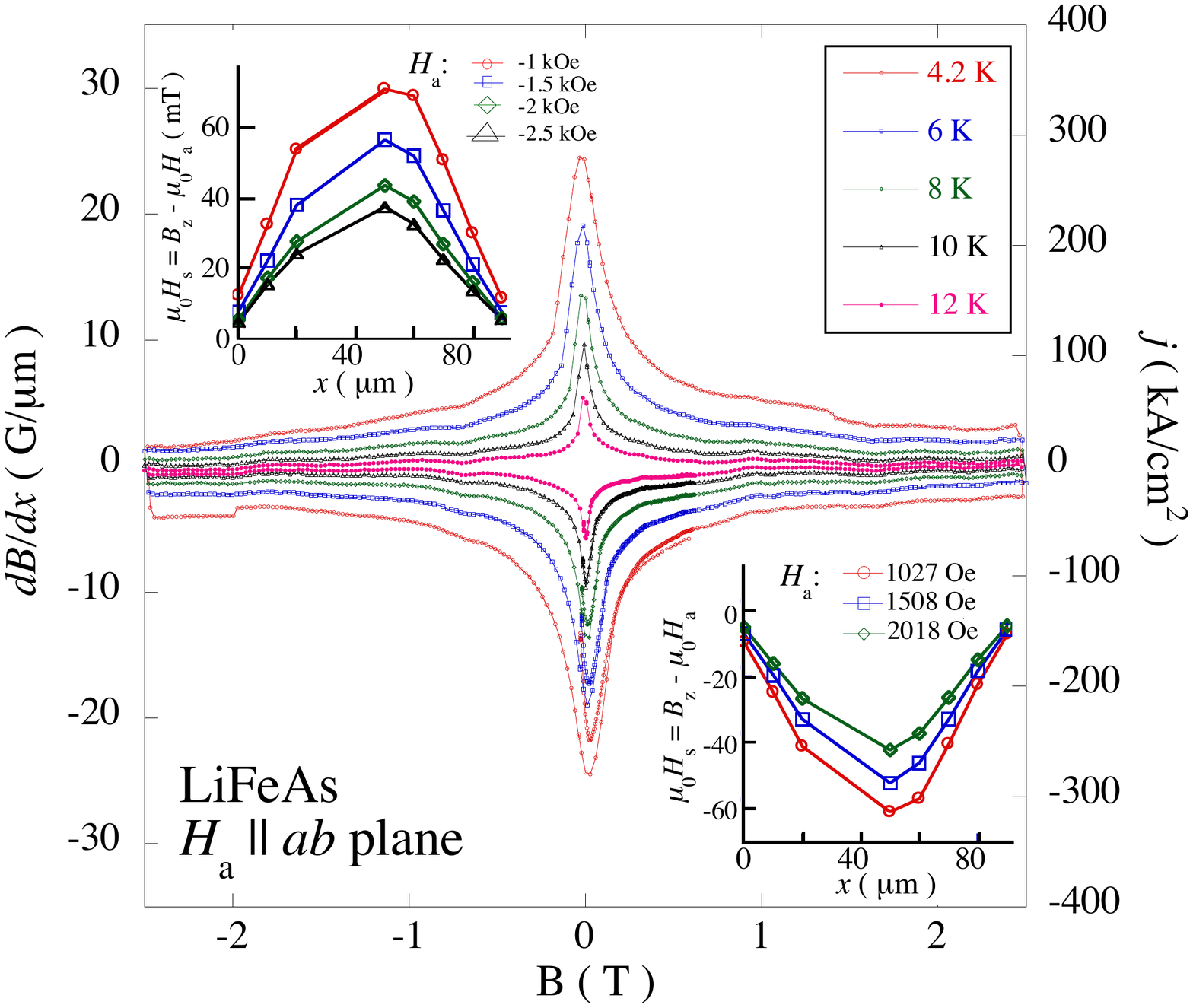}%
\vspace{-7mm}
\caption{(Color online)  Main panel: Hysteresis loops of $dB/dx \parallel ab$, for ${\bf B} \parallel  ab$, after zero field--cooling, measured at 4.2, 6, 8, 10, and 12 K. The right-hand ordinate shows the value of the corresponding current density $j_{c}^{ab}$. Upper inset: Profiles of the sample ``self--field'' $B - \mu_{0}H_{a}$ on the decreasing field branch (third quadrant), at various $H_{a}$--values. Lower inset: Profiles of the  ``self--field''  on the increasing field branch (first quadrant), at various $H_{a}$--values. }
\label{fig2}%
\end{figure}

%\section{Results}
The top inset in Fig.~\ref{fig1} shows typical profiles of $B$ measured after cooling in zero magnetic field (ZFC), application of a external field $\mu_{0}H_{a} = 2$~T $\parallel c$, and warming. The straight-line profiles are quite regular and conform to the Bean model,\cite{Bean62,Zeldov94} which implies a homogeneous critical current density that is practically field-independent over the range of $B$--values in the crystal. %due to the screening current is about 150 G, much less than the value of $H_{a}$). 
To obtain the local screening current, we plot the spatial gradient $dB/dx$ versus $B$. The main panel in Fig.~\ref{fig1} shows representative hysteresis loops of $dB/dx$ measured at 4.2, 8 and 12 K. The right ordinate shows the value of the corresponding current density $j_{ab}^{c} = (2/\mu_{0}) dB/dx$. 
%The factor of 2 takes into account the reduction of the magnetic field on the top of a semi-infinite crystal. 
The factor 2 corresponds to the case when $B$  is measured on the end surface of a semi-infinite superconducting slab; a more precise evaluation can be done using the results of %finite-size formulae derived by 
Brandt.\cite{Brandt98} The $j_{ab}^{c}$--values, of the order of  $ 100$ kA/cm$^{2}$, are similar to those obtained from global measurements in the same configuration.\cite{Pramanik2010LiFeAs} 
%Note that the typical experimental time constant for the Hall-probe technique is much smaller than that of commercial magnetometers, so that the effect of flux creep is smaller and the estimated current is much closer to the critical current density.
Because of flux creep, the measured current densities are slightly reduced with respect to the ``true'' critical current density, by a multiplicative factor determined by the effective experimental time scale (here, about 3 s).\cite{vdBeek92} The creep rate is rather modest;\cite{Pramanik2010LiFeAs} in our experiment, it amounts to 2-4 \% per decade of time, and is similar for $j_{ab}^{ab}$ and $j_{c}^{ab}$, so that the ratio $j_{ab}^{ab}/j_{c}^{ab}$ we shall be interested in is not appreciably altered. 

The shape of the $dB/dx$-hysteresis loop is very similar to that obtained for other iron-based superconductors.\cite{Kees,Kees1} It is characterized by a sharp maximum of the critical current density for $|B| \lesssim 6$ kG, behavior characteristic of a dominant contribution from strong pinning\cite{Ovchinnikov91,vdBeek2002} by nm-sized inhomogeneities.\cite{Sultan}  The constant $dB/dx$ at higher fields comes from a weak ``collective''  pinning contribution\cite{Blatter94} due to scattering of quasiparticles in the vortex cores by atomic-scale point defects.\cite{Kees,Kees1}  Figure \ref{fig2} shows similar results for $H_{a} \parallel ab-$plane and the Hall array  $\perp c$, the configuration that probes $j_{c}^{ab}$. Again, the flux density profiles are very well described by the Bean model, although in this field orientation, the critical current density is dominated by the strong pinning contribution over the whole field range. Due to the elongated slab geometry, the configuration with $H_{a} \parallel ab$ does not involve a demagnetization correction, so that the relation $j_{ab} = (2/\mu_{0}) dB/dx$ is practically exact. With $j_{c}^{ab}$ and $j_{ab}^{ab}$ both measured in this orientation, geometry-related corrections play no role in the determination of  $j_{ab}^{ab}/j_{c}^{ab}$.

\begin{figure}[t]%
\vspace{-5mm}
\includegraphics[width=1.4\linewidth]{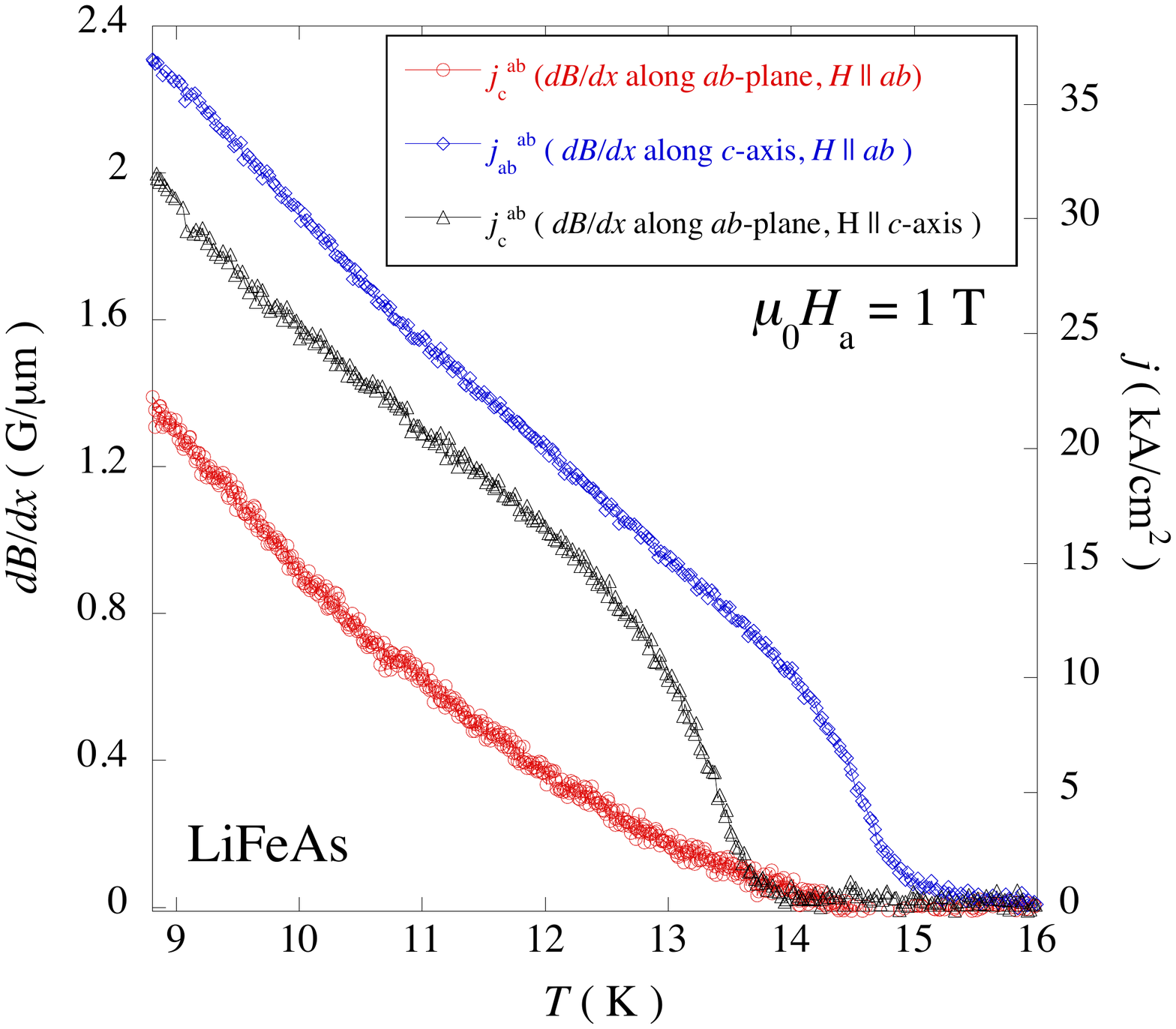}%
\vspace{-8mm}
\caption{(Color online) Local gradient of the magnetic induction measured in the three different configurations as function of temperature, for an applied field $\mu_{0}H_{a} = 1$T : (\color{red}$\circ$\color{black}) $dB/dx$ along $ab$ with ${\bf B} \parallel ab$, {\em i.e.}, $j_{c}^{ab}$;  (\color{blue}$\diamond$\color{black}) $dB/dx$ along $c$ with ${\bf B} \parallel ab$, {\em i.e.}, $j_{ab}^{ab}$;  ($\triangle$) $dB/dx$ along $c$ with ${\bf B} \parallel c$, {\em i.e.}, $j_{ab}^{c}$. }
\label{fig3}%
\end{figure}

The critical currents for the three directions are summarized in Fig.~\ref{fig3}, for an applied field of 1 T. Clearly, $j_{ab}^{ab}$ involving vortex motion along the $c-$axis (with vortices crossing the Fe-As planes) exceeds the other two critical currents. As expected,   $j_{c}^{ab}$ for easy vortex sliding along the $ab$--plane is the smallest. The critical current  $j_{ab}^{c}$ goes to zero at a lower temperature, reflecting the anisotropy of the irreversibility line in this material. 

\begin{figure}[t]%
\includegraphics[width=1.2\linewidth]{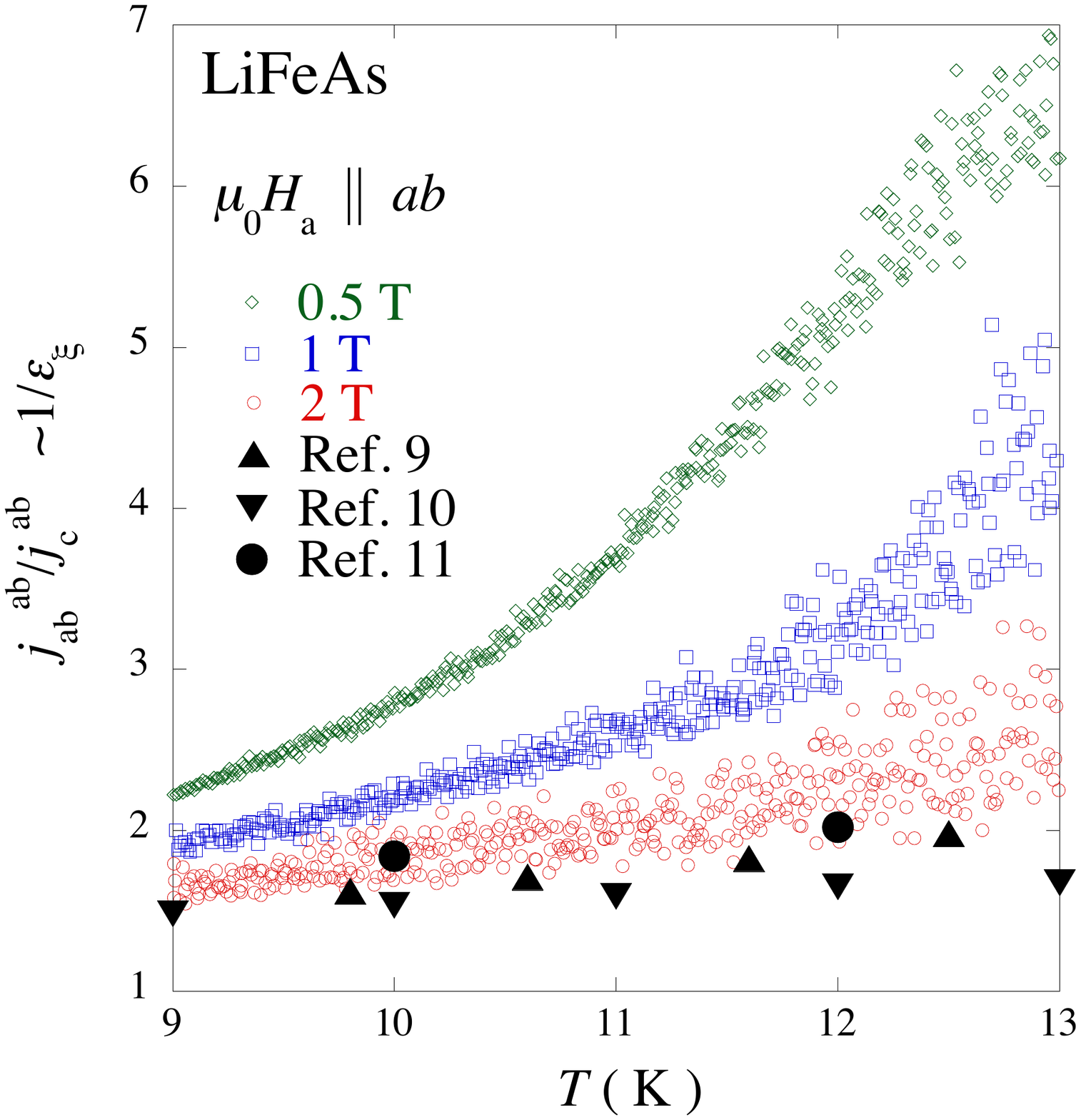}%
\vspace{-8mm}
\caption{(Color online) Critical current ratio $j_{ab}^{ab}/j_{c}^{ab} \sim 1/\varepsilon_{\xi}$ for applied magnetic fields of (\color{green}$\Diamond$\color{black}) 0.5 T; (\color{blue}$\Box$\color{black}) 1 T; (\color{red}$\circ$\color{black}) 2 T.}%
\label{fig4}%
\end{figure}

The critical current ratio $j_{ab}^{ab}/j_{c}^{ab}$ for ${\bf B} \parallel ab$ is plotted in Fig.~\ref{fig4} for different values of the applied field. To analyze it, we first consider  
theoretical results %for the critical current anisotropy 
derived for the case of  weak collective pinning.\cite{Blatter94} More specifically, in the regime of field--independent ``single--vortex''  pinning, 
%the $ab$--plane critical current density is independent of the field angle, {\em i.e.} $j_{ab}^{c} =j_{ab}^{ab}$. The 
the softer tilt- and shear moduli for vortex motion within the $ab$--plane imply that $j_{c}^{ab} = \varepsilon j_{ab}^{ab}$.\cite{Blatter94} This expression that does not take into account possible differences between $\varepsilon_{\lambda}$ and $\varepsilon_{\xi}$. %Rederiving the expressions of Ref.~\onlinecite{Blatter94}  but keeping track
A rederivation that keeps of the different contributions to the anisotropy yields  $j_{c}^{ab}  =  (\varepsilon_{\lambda}^{5/3}/\varepsilon_{\xi}^{2/3} ) j_{ab}^{c}$ and $j_{ab}^{ab} =(\varepsilon_{\lambda}/\varepsilon_{\xi})^{7/3} j_{ab}^{c}$. Hence, the anisotropy ratio
\begin{equation}
j_{ab}^{ab}/j_{c}^{ab}  =  \varepsilon_{\lambda}^{2/3} / \varepsilon_{\xi}^{5/3}
\end{equation}
is mainly determined %by that of the vortex core, {\em i.e.} 
by the coherence length anisotropy.

In the present situation though, the strong pinning contribution dominates the critical current density. Then,
%Given the fact that the strong pinning contribution dominates the critical current density for ${\bf B} \parallel ab$, we need to evaluate the effect of anisotropy for that case as well. In the strong pinning regime, 
the critical current density is determined by the direct sum of the elementary force $f_{p}$ that individual inhomogeneities exert on the vortex lines.\cite{Ovchinnikov91,vdBeek2002} It is given by the expression $j_{c} = (f_{p}/\Phi_{0}) n_{p}u_{0}^{2}$,\cite{vdBeek2002} where $n_{p}$ is the defect density, and $\Phi_{0}$ is the flux quantum. The trapping radius  $u_{0}$ is the largest distance, perpendicular to the field direction, on which a pin can be effective. The critical current anisotropy is thus determined by the anisotropy of $f_{p}$, and that of $u_{0}$.  The former is determined by the anisotropy of $\lambda$ and $\xi$, and by the geometric anisotropy of the pins, 
$\varepsilon_{b} =  \ln \left( 1 + b_{ab}^{2}/2 \xi_{ab}^{2} \right) / \ln \left( 1 + b_{ab}b_{c}/2\varepsilon_{\xi} \xi_{ab}^{2} \right) < 1$. Here, $b_{ab}$ and $b_{c}$ are the mean extent of the pins in the $ab$ and $c$--direction, respectively. At low fields, the $u_{0}$--anisotropy is determined only that of the vortex line tension, and is therefore field-independent. We find that $j_{c}^{ab} =  \varepsilon_{\lambda}^{2}\varepsilon_{b}^{-3/2} j_{s}^{c}$, while $j_{ab}^{ab} =(\varepsilon_{\lambda}^{2}/\varepsilon_{b}^{3/2}\varepsilon_{\xi}) j_{s}^{c}$. At higher fields, $u_{0}$ is determined by the intervortex interaction, leading to the ubiquitous decrease of the critical current density as $B^{-1/2}$. Then, $j_{c}^{ab} =  \varepsilon_{b}^{-2} \varepsilon_{\lambda}  j_{s}^{c}$, while $j_{ab}^{ab} =     (\varepsilon_{\lambda}/ \varepsilon_{b}^{2}\varepsilon_{\xi})  j_{s}^{c}$. In both cases,
\begin{equation}
j_{ab}^{ab}/j_{c}^{ab}  =  1/ \varepsilon_{\xi}.
\end{equation}
Thus, the experimental ratio $j_{ab}^{ab}/j_{c}^{ab}$, plotted in Fig.~\ref{fig4}, directly measures the coherence length anisotropy.

In spite of the fact that we could only evaluate the anisotropy above $T = 9$~K, it is clear that the extrapolated values of $1/\varepsilon_{\xi}$ at low temperature are of the order 1.5 -- 2. The anisotropy ($\sim 1/\varepsilon_{\xi}$) increases with increasing temperature to become as large as 6--7 at $T = 13$ K, an experimental upper limit imposed by the increasing role of flux creep at higher $T$. The anisotropy becomes smaller and less $T$-dependent at higher  magnetic field, and merges with the results obtained from the $B_{c2}$-ratios reported in Refs.~\onlinecite{Cho2011_LiFeAs_Hc2,Kurita2011Hc2,Khim2011} for a field as low as 2~T. Both the magnitude and the $T$-dependence of $\varepsilon_{\xi}$ are reminiscent of that of $\varepsilon_{\lambda}$ obtained on the 1111 family of iron--based superconductors.\cite{Okazaki2009} Notably, $\varepsilon_{\xi}$ is strongly temperature dependent at low fields, and less so at higher magnetic fields. 

Since the Fermi velocity is unaffected by field, a plausible framework for our observations is the temperature- \cite{Komendova2011} and field-dependent relative contribution of the two superconducting gaps to the effective superconducting coherence length. In particular, the evolution of $\varepsilon_{\xi}$ suggests that the relative weight of the gap on the more two-dimensional hole surface progressively decreases as the magnetic field is increased. For fields higher than 2~T, the gap on the three-dimensional electron surface would determine all superconducting properties related to the coherence length. This is consistent with recent thermal conductivity measurements that suggest that at fields as low as $0.1 B_{c2}(0)$ ({\em i.e.} 2~T), LiFeAs behaves as a single band superconductor. In that limit, the anisotropy of the coherence length and of the penetration depth are expected to be similar, and rather temperature independent. This is indeed the trend observed in the measurements: the high-field coherence length anisotropy seem to behave very similarly to reported results for the penetration depth anisotropy.\cite{Sasmal2010} It is to be noted that as the magnetic field is increased, the vortex core radius should plausibly shrink,  such as this occurs in NbSe$_{2}$.\cite{Sonier2005} Also, the core structure should be modified.\cite{Komendova2011} This does not affect the ratio of the coherence lengths discussed here. 

The field-dependence of $\varepsilon_{\xi}$ may explain why the weak collective pinning contribution to the critical current density is more important for fields oriented parallel to $c$. The values of $\varepsilon_{\xi}$ and $\varepsilon_{\lambda}$ are very similar at  fields above 1 -- 2 T at which this contribution manifests itself. Hence, the weak pinning part of the critical current should be nearly the same for the two field orientations, as in a single band superconductor. At lower fields, it should be enhanced for $H \parallel ab$, but this is not perceptible because it remains masked by the strong pinning contribution. On the other hand, strong pinning is enhanced for all values of $H \parallel ab$  because its dependence on $\varepsilon_{\xi}$ through  $\varepsilon_{b}$ .

In conclusion, we present a direct technique for the measurement of the critical current anisotropy in uniaxial type II superconductors. The technique crucially relies on the use of a local probe of the magnetic induction, in this case, miniature Hall probe arrays. In the situation of strong pinning by extrinsic extended point defects, the ratio of the critical current densities along the $ab$--plane and the $c$-axis, for field oriented along the $ab$-plane, directly yields the coherence length anisotropy. We apply the method to infer the coherence length anisotropy $1/\varepsilon_{\xi}$ of LiFeAs at much lower magnetic fields than commonly reported. We interpret the results in terms of the gap anisotropy, and find that this is reduced to its value near $B_{c2}$ by the application of a magnetic field as low as 2 T .

\begin{acknowledgments}

We thank V.G. Kogan for useful discussions and Dr. S. Bansropun and his group at Thales-TRT, Palaiseau for the careful processing of the Hall sensors. This work was supported by the French National Research agency, under grant ANR-07-Blan-0368 ``Micromag''. The work at The Ames Laboratory was supported by the U.S. Department of Energy, Office of Basic Energy Sciences, Division of Materials Sciences and Engineering under contract No. DE-AC02-07CH11358. Work at SKKU was partially supported by Basic Science Research Program through the National Research Foundation of Korea (NRF) funded by the Ministry of Education, Science and Technology (2010-0007487). The work of R. Prozorov in Palaiseau was funded by the St. Gobain Chair of the Ecole Polytechnique. 
\end{acknowledgments}

\end{document}